\documentclass{test_aa}  

\usepackage{graphicx}
\usepackage{txfonts}

\begin{document}

\title{Influence of Departures from LTE on Determinations of the Sulfur Abundances in A–K Type Stars}

\author{S.~A.~Korotin
        \inst{1}
        \and
        K.~O.~Kiselev
        \inst{1}
}

   \institute{Crimean Astrophysical Observatory, Nauchny, Crimea, Russia\\
              \email{serkor1@mail.ru}
}

\date{Received date; accepted: date}

\abstract
{The influence of departures from local thermodynamic equilibrium (LTE) on 
neutral sulfur lines is considered. A grid of corrections is proposed to take 
into account the influence of departures from LTE for neutral sulfur lines in 
the visible and infrared spectral regions, including the H-band. The grid is 
calculated using the atomic model of sulfur incorporating the most up-to-date 
collision rates with electrons and hydrogen. The inclusion of levels and 
transitions of ionized sulfur in the atomic model made it possible to expand 
the range of effective temperatures of stellar photospheres in the grid up to 
10000\,K. The atomic model was tested in determining the sulfur abundance of 
13 stars and showed its adequacy in a wide range of fundamental stellar 
parameters. In the spectra of all test stars, the sulfur lines are fitted with 
similar abundances of the element, regardless of the degree of influence of the
effects of deviation from LTE on a particular spectral line. For lines of 
several multiplets, the wavelengths and oscillator strengths were refined. A 
list of S\,I lines recommended for determining sulfur abundance has been 
created.
}


\authorrunning{Korotin and Kiselev}
\titlerunning{NLTE-effects of the Sulfur Abundances in A–K Type Stars}

\maketitle

\section{INTRODUCTION}
To understand the processes of chemical evolution in the Galaxy and the history
of star formation, researchers examine the relationships between various 
chemical elements in stellar atmospheres. Iron-peak elements and $\alpha$-elements 
play a fundamental role in these studies, as their nucleosynthesis occurs over 
different timescales. $\alpha$-Elements can be used as cosmic clocks to monitor both 
stellar nucleosynthesis and the history of galaxy formation. Unlike other 
$\alpha$-elements, sulfur is moderately volatile. For this reason, its abundance 
measured in stars within the Local Group galaxies can be directly compared to 
its abundance measured in extragalactic H II regions. While $\alpha$-elements are 
produced by Type II supernovae, iron-peak elements are produced by Type Ia 
supernovae over longer timescales. Among the various $\alpha$-elements, the 
nucleosynthesis of sulfur is not fully understood. Sulfur is formed during the 
final stages of evolution of massive stars (M~>~20 M$_\odot$). Hydrostatic burning of 
neon in the core at temperature of $1.2\times10^9$~K leads to the formation of a 
convective oxygen core and the production of $\alpha$-elements, including sulfur. 
Further, various nuclear transformation processes occur, resulting in the 
near-total destruction of the produced sulfur during the Si-burning phase. 
However, sulfur production continues with oxygen burning in the outer layers 
around the core, as well as through explosive oxygen burning during Type II 
supernova explosions \citep{Limongi03,Kobayashi20}. This enrichment is assumed to occur over a 
relatively short timescale on the order of tens of millions of years. 
Accordingly, it can be expected that the [S/Fe] ratio will be relatively 
constant for low-metallicity stars. These are stars that formed before the 
ignition of the first Type Ia supernovae. This ratio then decreases as iron 
levels increase. There is still no consensus on whether this “plateau” exists 
in the relative sulfur abundance of metal-deficient stars. The authors of \cite{Israelian01} 
suggested that the [S/Fe] ratio continues to increase at [Fe/H]~<~2.0. Other 
studies ((\cite{Spite11}, \cite{Kacharov15}, and others) have shown evidence of a plateau, although this 
plateau has significant dispersion due to the complicated analysis of weak 
lines at such low metallicities. In stars with near-solar heavy element 
abundance, sulfur behaves as a typical $\alpha$-element. Within the range of 
-1~<~[Fe/H]~<~0, the [S/Fe] ratio decreases to zero, which confirms the general 
characteristics of $\alpha$-element nucleosynthesis.

However, sulfur has received little attention from researchers. Analysis of 
sulfur is often omitted in favor of other $\alpha$-elements, whose measurements are 
less complex. The issues at hand include the weakness of sulfur’s optical 
lines. In the optical range, there are 6, 8, and 10 multiplets with relatively 
weak lines that nearly vanish in metal-deficient stars. Two triplets in the
near-infrared  region, $\lambda$\,9212–9237 and $\lambda$\,10455–10459 \AA\ are heavily affected by 
departures from local thermodynamic equilibrium (LTE) (\cite{Takeda05}, \cite{Korotin09}). These lines are 
located in a region of the spectrum that is significantly distorted by lines 
from Earth’s atmosphere. For these reasons, studies of sulfur are often 
bypassed in favor of simpler analyses of other $\alpha$-elements. As a result, our 
understanding of sulfur’s behavior is still insufficient compared to other 
$\alpha$-elements.
 
The influence of non-LTE effects on sulfur lines requires detailed study. 
As shown in \cite{Takeda05} and \cite{Korotin09}, non-LTE corrections for S\,I lines have complex dependences 
on the fundamental parameters of stars and the sulfur abundance itself. Model 
calculations \cite{Takeda05} and \cite{Korotin09} show similar patterns in the behavior of non-LTE 
corrections depending on stellar parameters, though the corrections differ 
slightly in magnitude. It should be noted that since these model calculations 
were conducted, new atomic data have become available for neutral and ionized 
sulfur. The use of new detailed quantum- mechanical calculations of inelastic 
collision rates with electrons and hydrogen should help avoid potential 
errors introduced by approximate formulas. We have worked on updating the 
sulfur atomic model \cite{Korotin09}, incorporating the latest atomic data and verifying 
it with spectra of well-studied stars.

\section{SULFUR ATOM MODEL}

Our previously developed sulfur atom model \citep{Korotin09} included 64 levels 
of neutral sulfur and the ground level of S\,II. Approximate formulas were used 
to account for collisional interactions with electrons and hydrogen, though they 
were often of limited accuracy. Over time, our knowledge of the atomic 
parameters of sulfur has significantly improved. Detailed quantum-mechanical 
calculations have occurred for collisional rates of inelastic interactions 
between 56 levels of neutral sulfur and hydrogen \citep{Belyaev20}. Previously,
the collisions with hydrogen in the atomic model were accounted for using the 
so-called “Drawin formula” \citep{Drawin68, Steenbock84}. It is well known that 
this formula gives highly inaccurate results, which researchers attempted to 
correct by introducing empirical correction factors, obtained by adjusting 
non-LTE calculations to observations. Comparisons between this approximate 
formula and the detailed calculations \citep{Belyaev20} show discrepancies 
that can reach several orders of magnitude. A similar situation exists for 
collisional rates with electrons. The ADAS database \citep{Summers11} now includes 
detailed calculations for collisional interactions of the 17 lowest levels of 
sulfur with free electrons, allowing us to abandon the use of inaccurate 
approximate formulas. Moreover, applying the old atomic model to study sulfur 
lines in A-type stars with sufficiently high effective temperatures raised the 
need to include additional levels of S\,II and S\,III in the model. All this 
prompted us to upgrade the atomic model and perform the appropriate testing.

\begin{figure*}
\includegraphics[scale=0.50]{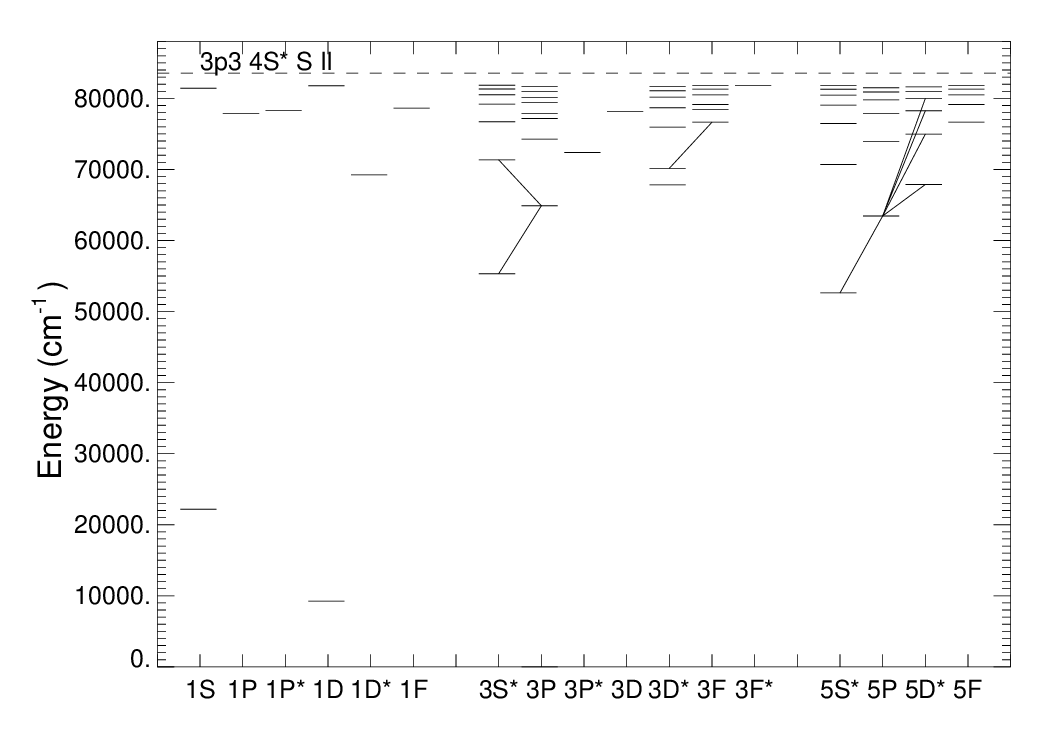}
\includegraphics[scale=0.50]{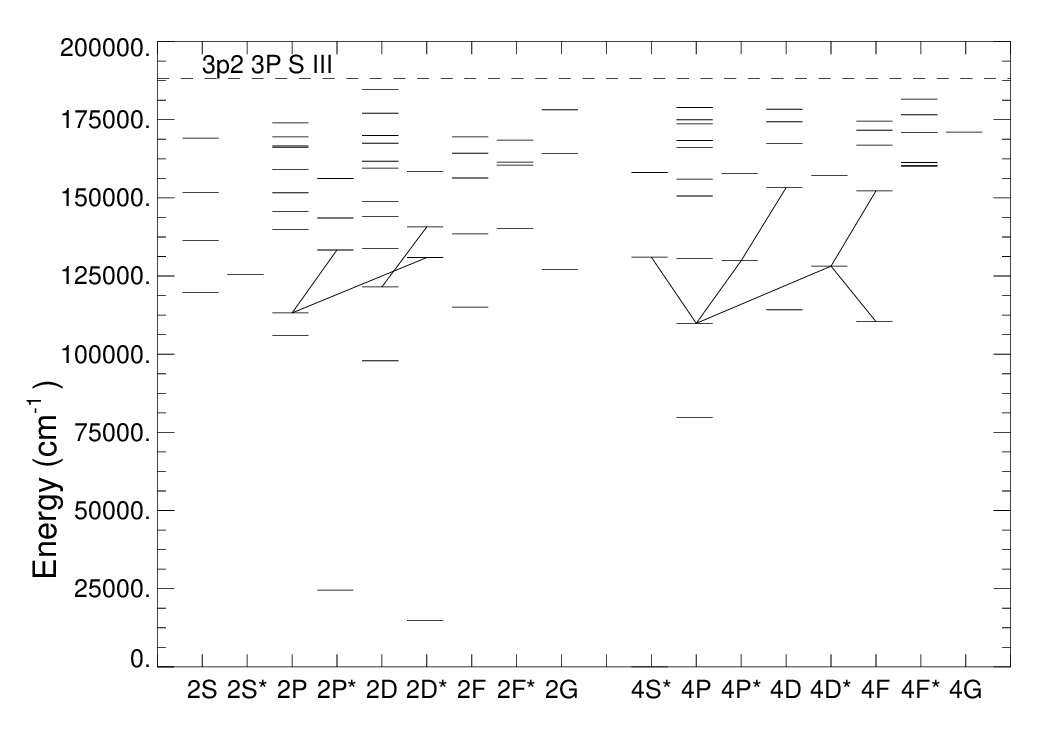}
\caption{Grotrian diagrams for S~I (left) and S~II (right). Only transitions used to determine sulfur abundances are shown.}
\label{Grot}
\end{figure*}

As in the previous model, the populations of 64 energy levels of neutral sulfur 
were considered in detail. The highest S\,I level in the model is separated 
from the continuum by 0.21\,eV, which ensures reliable interaction at 
temperatures above 2500\,K. Additionally, 81 levels of S\,II and the ground 
level of S\,III were included. The highest S\,II level is separated from the 
continuum by 0.45\,eV, corresponding to a temperature of 5150\,K. To more fully 
account for particle numbers, we included one level of S\,I, six levels of 
S\,III, and the ground level of S\,VI, with populations calculated in LTE. 
Fine structure levels were not considered. The values of electronic energy 
levels were taken from \cite{Martin90}. The diagram of the sulfur atom levels 
is shown in Fig.~\ref{Grot}.

Detailed consideration included 775 bound-bound and 146 bound-free transitions 
between levels. Line profiles were calculated using the Voigt profile, taking 
into account radiative broadening as well as Stark and Van der Waals broadening. 
The number of points along the profile varied from 30 to 110, depending on the 
line’s intensity. Photoionization cross sections and oscillator strengths were 
taken from the TOPBase database \citep{Cunto93}. For forbidden transitions, oscillator 
strengths were obtained from the catalog \cite{Hirata94}.

Detailed quantum-mechanical calculations of collisional excitation rates by 
electrons for transitions among 17 levels of neutral sulfur were obtained from 
the ADAS database \citep{Summers11}. For transitions among the 32 lower levels 
of ionized sulfur, we used calculations from \cite{Tayal10}. For other allowed 
transitions, we applied the formula from \cite{Regemorter62}, and for forbidden 
transitions, the formula from \cite{Allen73}, with an effective collision 
strength set to 1. Collisional ionization was accounted for using the formula 
from \cite{Seaton62} with the photoionization threshold cross-section from 
TOPBase \citep{Cunto93}.

Collisional interactions with hydrogen atoms become very important in the 
atmospheres of stars with low effective temperatures and metal-poor stars, 
where electron density decreases. Although the actual collisional rates with 
hydrogen are quite small, the high concentration of hydrogen provides a 
significant influence on the redistribution of electronic level populations. 
Recently, detailed quantum-mechanical calculations for inelastic collisions 
between various atoms and hydrogen have made it possible to account for this 
effect more realistically than with approximate approaches. For the sulfur 
atom, such calculations were performed in \citep{Belyaev20}. We incorporated 
collisional interactions with hydrogen for forty levels of neutral sulfur 
within a temperature range of 1000 to 10000\,K. Beyond this range, 
extrapolation was applied; however, the effect of collisional interactions 
with hydrogen becomes negligible at such high temperatures.

To obtain the sulfur level populations in stellar atmospheres, we used the 
MULTI software package \citep{Carlsson86}, version 2.3, which we modified 
slightly \citep{Korotin99}. The main changes involve using the complete opacity 
calculation package from the ATLAS9 software \citep{Castelli03}, including the 
so-called opacity distribution function (ODF) arrays. After solving the 
statistical equilibrium and radiative transfer equations together, we obtained 
coefficients of deviation from LTE populations, known as b-factors: the ratio 
of non-LTE to LTE populations ($N_{\rm NLTE}/N_{\rm LTE}$). These were passed 
to the synthetic spectrum calculation software SynthV \citep{Tsymbal19}, where 
sulfur line profiles were calculated with non-LTE effects, while lines of other 
elements were calculated under the LTE approach. Line parameters necessary for 
synthetic spectrum calculations were taken from the VALD database 
\citep{Ryabchikova15}.

\section{APPLICATION OF THE SULFUR ATOM MODEL TO SOLAR SPECTRUM ANALYSIS. 
REFINEMENT OF NEUTRAL SULFUR LINE PARAMETERS}

After updating the sulfur atom model, it is necessary to validate its accuracy. 
This requires comparing the model with observed spectra from several stars with 
well-known atmospheric parameters. These stars’ atmospheric parameters should 
span a broad range of temperatures and pressures. Validation of the model is 
performed by comparing observed and synthetic profiles of spectral lines that 
belong to different multiplets and exhibit different degrees of deviation from 
values predicted in the LTE approximation. For instance, some spectral lines 
show no considerable non-LTE effects, while others display significant 
deviations from LTE. Such a combination of lines cannot be accurately modeled 
in an LTE approach by simply adjusting the element’s abundance. A well-selected 
atomic model in non-LTE calculations should allow an acceptable description of 
this spectrum using a consistent element abundance for all analyzed lines.

Only a limited number of S\,I lines are suitable for determining sulfur 
abundance in the visible spectrum. These include the first 
($\lambda$\,9212–9237~\AA), sixth ($\lambda$\,8694~\AA), eighth 
($\lambda$\,6743-6757~\AA), and tenth ($\lambda$\,6046-6052~\AA) quintet system 
multiplets. In the near-infrared range, multiplet lines at 
$\lambda$\,10455-10459~\AA, $\lambda$\,15400-15422~\AA, 
$\lambda$\,15469-15478~\AA\ and $\lambda$\,22507-22707~\AA. The remaining lines 
are either too weak or blended. Notably, the eighth and tenth multiplets each 
consist of three lines, and each line itself is a superposition of three 
components with slight wavelength shifts. Such line profiles deviate 
significantly from a Gaussian shape, necessitating synthetic spectrum 
calculations for comparison with observations, as illustrated in the figures 
below. Using equivalent widths for these lines requires caution and the 
application of methods for calculating equivalent widths of multicomponent 
lines.

For analyzing neutral sulfur lines, we used the solar spectrum from atlases 
\cite{Kurucz84} and \cite{Reiners16}, covering the range from $\lambda$\,3000 
to $\lambda$\,13000~\AA\ and $\lambda$\,4050 to $\lambda$\,23000~\AA, 
respectively. Calculations were based on a solar atmosphere model 
\cite{Castelli03} with a microturbulent velocity of 1 km/s. The Sun’s 
rotational velocity was set at 1.8 km/s. Element abundances in the solar 
atmosphere were taken from \cite{Asplund21}. Profiles of all lines included 
Van der Waals broadening, calculated according to \cite{Barklem97,Barklem98}.

\begin{figure*}
\includegraphics[scale=0.30]{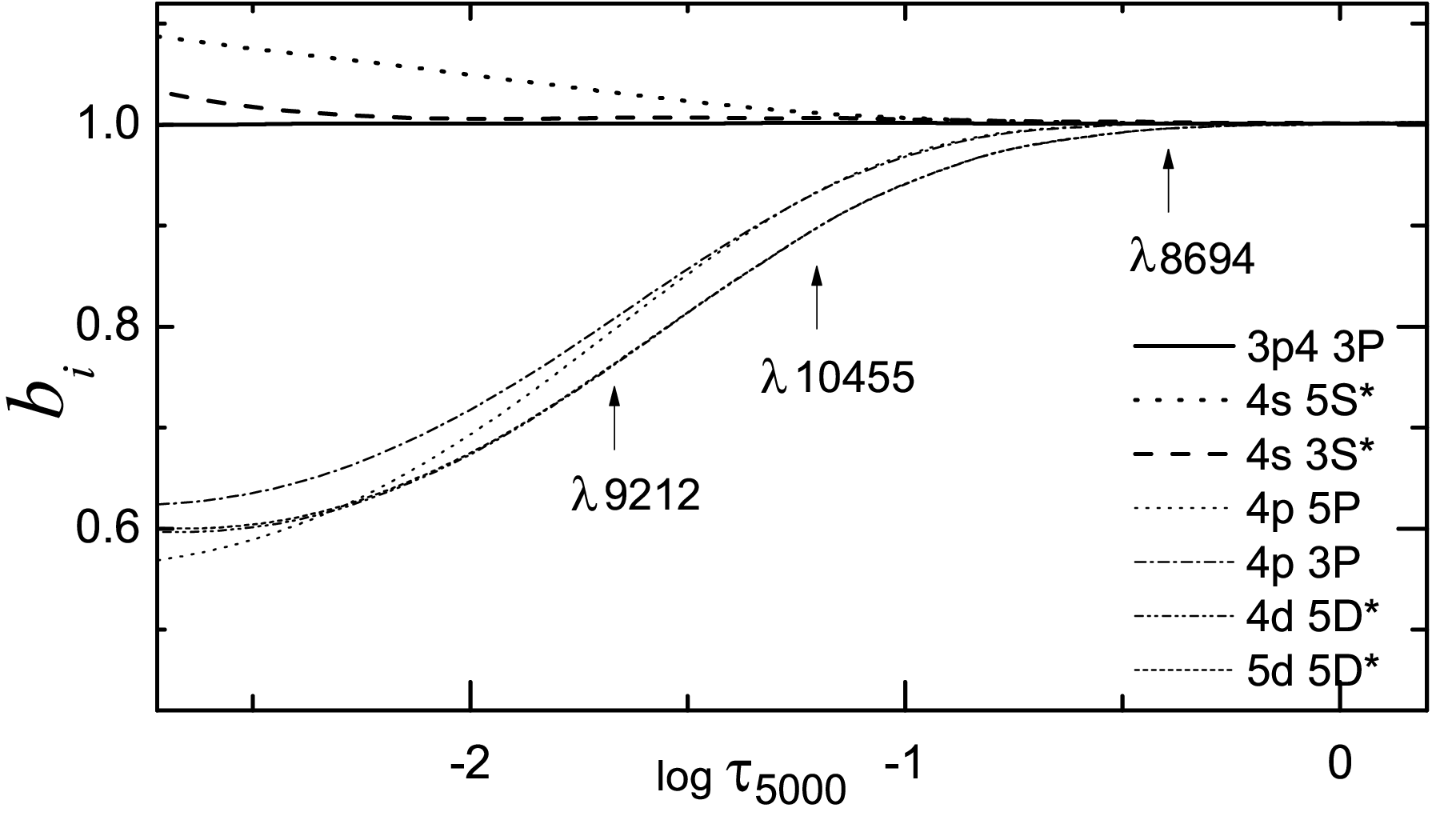}
\includegraphics[scale=0.30]{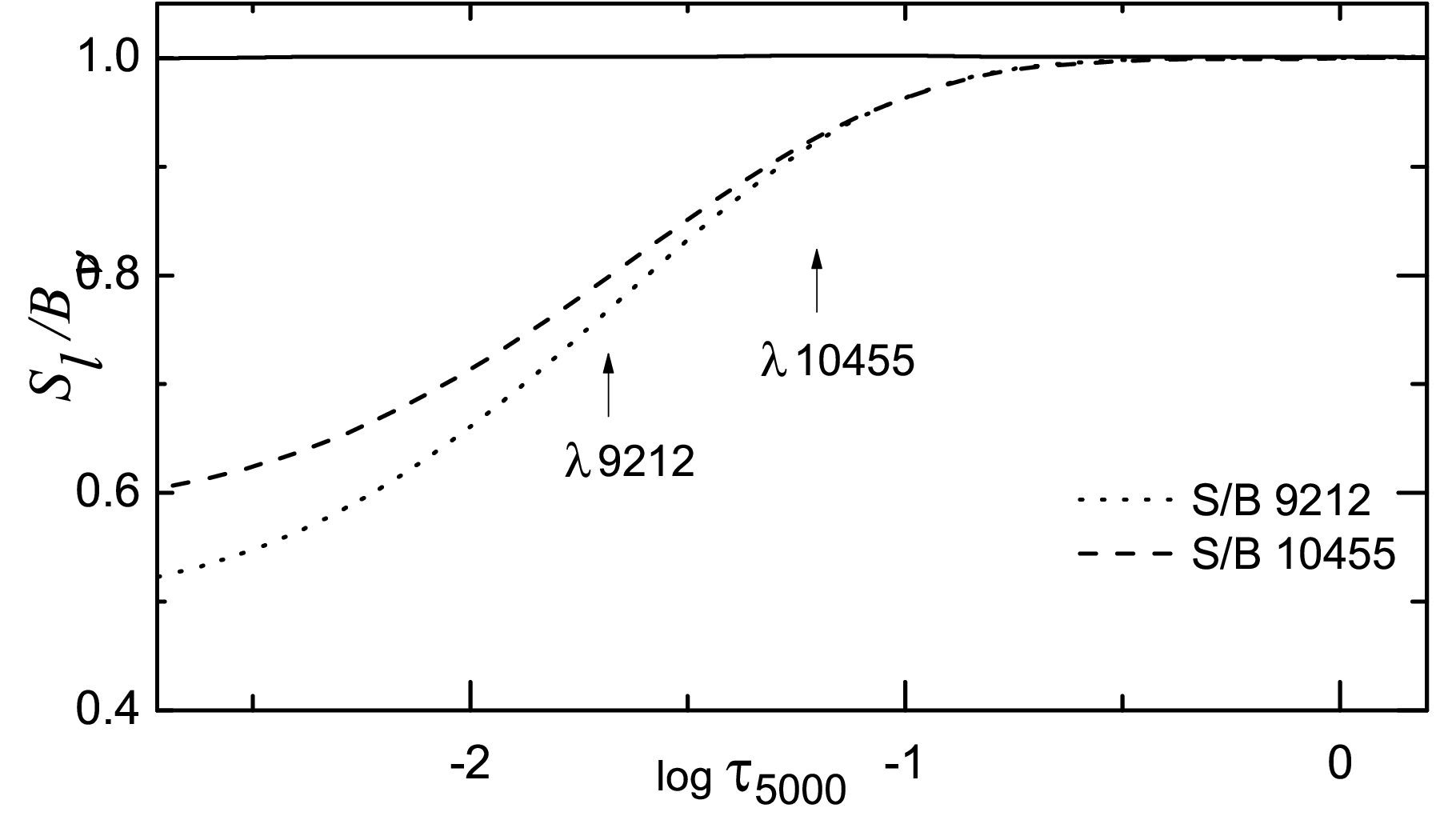}
\caption{Distribution of b-factors in the Sun’s atmosphere (left panel) and the 
variation of the S$_{l}$/B$_{\nu}$ ratio with depth for two IR triplets (right 
panel). Arrows indicate the formation depths of some sulfur lines}
\label{b_sun}
\end{figure*}

Calculations of neutral sulfur level populations for the solar photosphere 
showed results consistent with \cite{Korotin09}. The sixth, eighth, and tenth 
multiplet lines form almost in LTE and are not influenced by non-LTE effects. 
The lines in the IR H-range behave similarly. Only the lines of two IR triplets, 
$\lambda$\,9212-9237 and $\lambda$\,10455-10459~\AA\, are significantly 
enhanced due to overpopulation of their lower levels, 4s~5S* and 4s~3S*, and 
underpopulation of their upper levels, 4p~5P and 4p~3P, respectively, at the 
line formation depths. The b-factor distribution of some sulfur levels by 
photospheric depth is shown on the left panel of Fig.~\ref{b_sun}. The 
population levels between which the sixth, eighth, and tenth multiplet 
transitions form are in LTE at these line formation depths. The right panel of 
Fig.~\ref{b_sun} shows the ratio of the source function S$_{l}$ to the Planck
function B$_{\nu}$ for both IR triplets as a function of optical depth. It is 
evident that at the line formation depths, S$_{l}$/B$_{\nu}$~<~1. Thus, non-LTE 
effects strengthen the lines, leading to negative corrections to sulfur 
abundance estimates when using the LTE approach for these multiplet lines.

Experimentally, oscillator strengths and wavelengths have only been determined 
for the triplet lines at $\lambda$\,10455-10459~\AA\ \citep{Zerne97}. For other 
sulfur lines, only theoretical calculations are available, notably in 
\cite{Biemont93} (hereafter BQZ). More recent results are presented in
\cite{Zatsarinny06} (hereafter ZB). Studies \cite{Froese06} (hereafter FTI) and
\cite{Deb08} (hereafter DH) cover only part of the S\,I lines examined here.

Oscillator strengths for the lines of the first multiplet are nearly identical 
in ZB, FTI, and DH, with differences not exceeding 0.03 dex and an estimated 
accuracy of 25\%. For further calculations, we used data from ZB for these 
lines.

\begin{figure*}
\resizebox{\hsize}{!}{\includegraphics{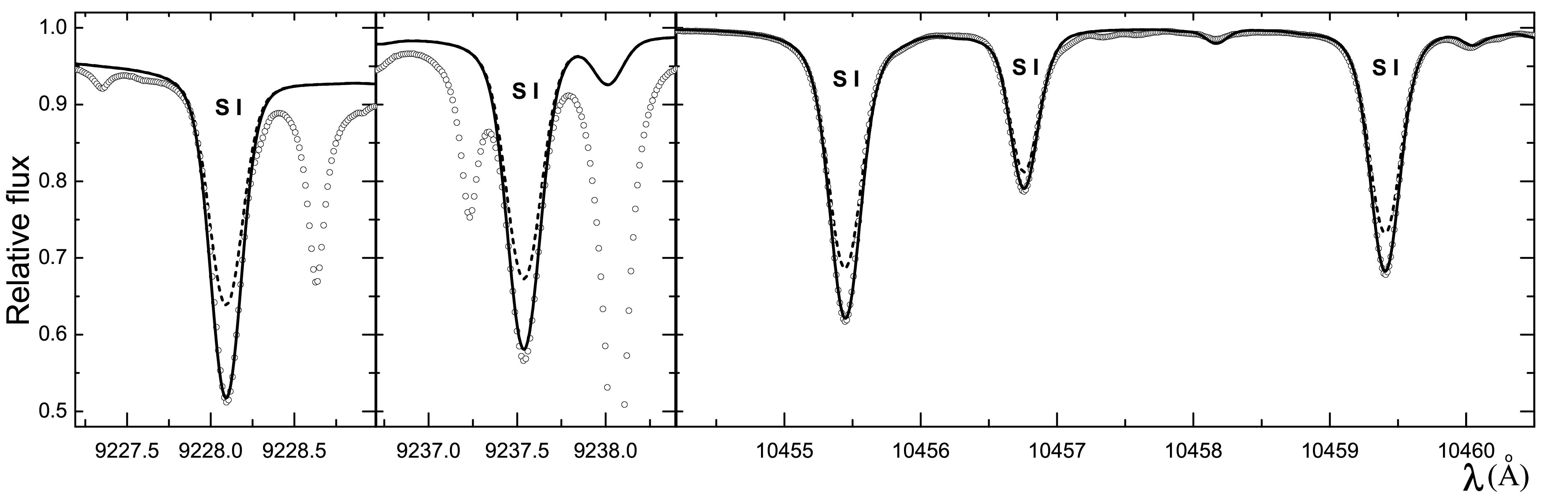}}
\caption{Comparison of observed (circles) and synthetic profiles of IR triplet 
lines in the Sun’s spectrum. The non-LTE profile for sulfur abundance 
(S/H) = 7.16 is shown as a solid line, and the LTE profile, calculated with the 
same abundance, is shown as a dashed line.}
\label{IR_prof}
\end{figure*}

Figure \ref{IR_prof} compares the calculated line profiles of two IR triplets, 
including non-LTE effects, with the observed solar spectrum. All lines except 
for $\lambda$\,9212~\AA, which is distorted by Earth’s atmosphere, are 
described well at a sulfur abundance (S/H )=7.16 , with a profile fitting 
error of no more than $\pm 0.02$ dex. The non-LTE correction for the lines of 
the first IR triplet ($\lambda$\,9212-9237~\AA) is –0.23 dex, while for the 
$\lambda$\,10456~\AA\ line, the non-LTE correction is –0.18 dex. This derived 
sulfur abundance aligns with the sulfur abundance in chondrites: 
(S/H)$=7.15\pm 0.02$ \citep{Lodders21}. The sulfur abundance in these 
meteorites is determined with high accuracy and can be used as a reference for 
the Solar System. Given that the Sun is a main-sequence dwarf star, its 
photosphere likely retains its original composition, so the sulfur abundance 
should match the meteoritic value. Agreement between the IR triplet-derived 
sulfur abundance and meteoritic levels can serve as a confirmation of the 
adequacy of our sulfur atom model.

The analysis of other sulfur lines in the optical range showed that using 
wavelengths and oscillator strengths from the VALD \citep{Ryabchikova15} or 
NIST \citep{Kramida12} databases to calculate the synthetic spectrum did not 
yield ideal line profiles. This is understandable, since the oscillator 
strength calculation accuracy varies from 25 to 40\% depending on the source. 
For example, for the sixth multiplet lines, the log gf values from ZB and BQZ
differ by 0.31 dex. Wavelengths are also calculated based on electronic level 
energies, which are typically determined with an accuracy of 
$\pm$25-30 cm$^{-1}$ \citep{Martin90}. Given that optical-range S\,I lines are 
unaffected by non-LTE effects, we refined their wavelengths and oscillator 
strengths. This procedure of refinement of line parameters serves for finding a 
balance between observational data and theoretical values of wavelengths and 
oscillator strengths.

The sixth multiplet is the triplet at $\lambda$\,8693-8694~\AA. The log\,gf 
values of this triplet from BQZ are systematically lower than ZB by 0.31 dex 
and lower than DH by 0.45 dex. Profiles calculated using BQZ oscillator 
strengths were much weaker than observed. ZB values produced a more realistic 
synthetic spectrum, though profiles were still slightly weaker than observed. 
At the same time, oscillator strengths from DH resulted in excessively strong 
calculated lines. The difference between log\,gf from DH and ZB is 0.13 dex for 
all components of the multiplet. The best agreement between synthetic and 
observed profiles was achieved with values of log\,gf$_{\rm ZB} + 0.088$. We 
also adjusted the multiplet component wavelengths. The shift does not exceed 
0.07~\AA\ from those in VALD, which is within the stated accuracy range. 

The eighth multiplet is the triplet at $\lambda$\,6743-6757~\AA. Each component 
in this triplet is itself a triplet of closely spaced lines. The log\,gf values 
of these nine lines from BQZ are on average lower than ZB by 0.11 dex. 
DH calculations for this multiplet are unavailable. When using BQZ oscillator 
strengths with meteoritic sulfur abundance, the theoretical lines are
weaker than observed, while ZB data produced slightly stronger lines than 
observed. The best agreement is achieved at log log\,gf$_{\rm ZB} + 0.075$, 
which is close to the BQZ values, but slightly lower. It should be noted that 
the wavelengths of these multiplet components vary by up to 0.1~\AA across 
different sources. The multiplet itself is a combination of transitions between 
three lower and five upper levels. Lower level energies are relatively 
well-determined, particularly for the upper levels of the first multiplet lines 
at $\lambda$\,9212-9237~\AA. By varying upper level energies, we attempted to 
achieve the best description of the observed profiles. Energy corrections did 
not exceed 25 cm$^{-1}$. The maximum wavelength shift relative to VALD 
recommendations was 0.06~\AA, which is consistent with various discrepancies 
from the literature. 
 
The tenth multiplet is the triplet at $\lambda$\,6041-6052~\AA. Each component 
in this triplet is also a triplet of closely spaced lines. The 
$\lambda$\,6041~\AA\ line is blended with a strong iron line and was excluded 
from analysis. DH and BQZ do not provide calculations for this multiplet, and 
VALD values are on average lower than ZB by 0.15 dex. Profiles calculated with 
these oscillator strengths were weaker than observed. A satisfactory description 
for the $\lambda$\,6052~\AA\ line was achieved by increasing log\,gf$_{\rm ZB}$
by 0.113 dex for all components in this triplet. We also adjusted the 
wavelengths of two line components, with a maximum shift of 0.03~\AA. However, 
this oscillator strength adjustment did not provide a good fit for the 
$\lambda$\,6046~\AA\ line in the observed spectrum. This line is clearly 
broadened by an unknown component in the solar spectrum. Thus, we excluded this 
line from further analysis.

\begin{figure*}
\resizebox{\hsize}{!}{\includegraphics{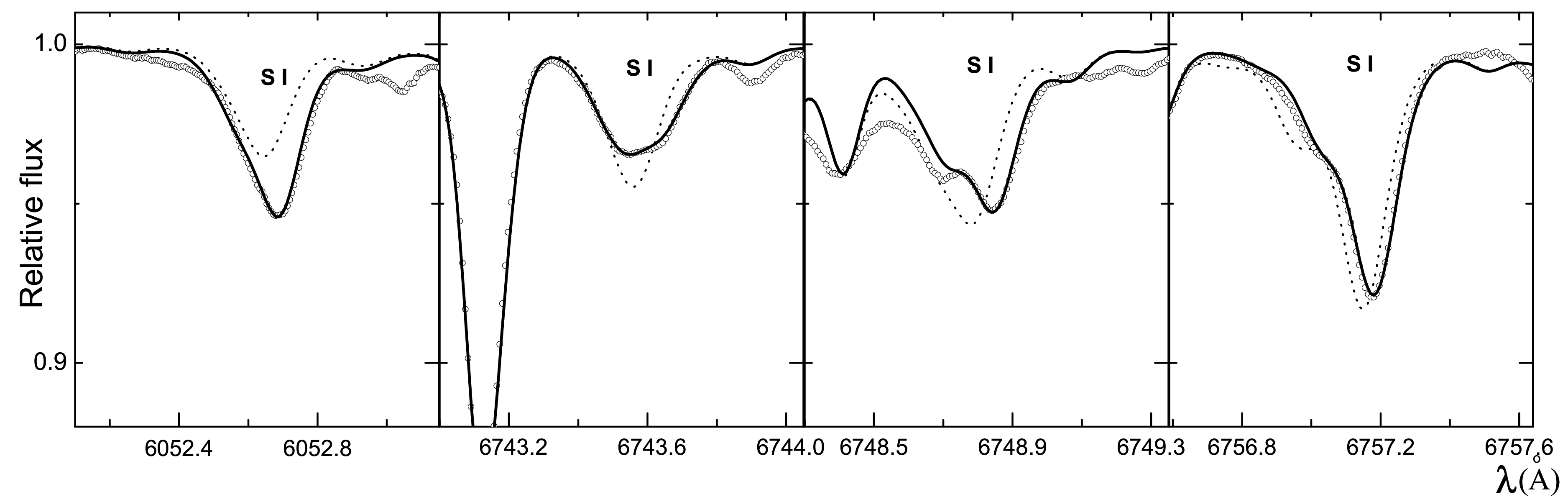}}
\caption{Comparison of observed (circles) and synthetic profiles of the eighth 
and tenth multiplet lines in the Sun’s spectrum. Profiles with refined 
parameters are shown as solid lines, and those calculated with VALD parameters 
are shown as dotted lines.}
\label{Sun_10_8}
\end{figure*}

Figure ~\ref{Sun_10_8} presents examples of synthetic profiles with refined 
parameters, describing the observed solar spectrum. The refined line parameters 
are listed in Table~\ref{lineSI}.

\begin{table}
\caption {Parameters of S I lines} 
\label{lineSI}
\tiny
\medskip
\begin{tabular}{cccc|cccc}
\hline
$\lambda$, \AA& E,eV& log\,gf&$\Gamma_{vw}$&$\lambda$,\AA& E,eV& log\,gf&$\Gamma_{vw}$\\
\hline
 6052.53&  7.87&  -1.99&  -6.55&10455.45&  6.86&   0.25&  -7.32\\
 6052.59&  7.87&  -1.15&  -6.55&10456.76&  6.86&  -0.45&  -7.32\\
 6052.69&  7.87&  -0.56&  -6.55&10459.41&  6.86&   0.03&  -7.32\\
 6743.47&  7.87&  -1.24&  -6.65&15400.08&  8.70&   0.43&  -7.00\\
 6743.54&  7.87&  -0.88&  -6.65&15403.72&  8.70&  -0.30&  -7.00\\
 6743.65&  7.87&  -0.94&  -6.65&15403.79&  8.70&   0.61&  -7.00\\
 6748.61&  7.87&  -1.36&  -6.65&15422.20&  8.70&  -1.84&  -7.00\\
 6748.71&  7.87&  -0.77&  -6.65&15422.26&  8.70&  -0.30&  -7.00\\
 6748.85&  7.87&  -0.56&  -6.65&15422.28&  8.70&   0.77&  -7.00\\
 6756.86&  7.87&  -1.71&  -6.65&15469.82&  8.05&  -0.17&  -6.90\\
 6757.03&  7.87&  -0.86&  -6.65&15478.48&  8.05&   0.06&  -6.90\\
 6757.18&  7.87&  -0.28&  -6.65&22507.56&  7.87&  -0.48&  -7.44\\
 8693.16&  7.87&  -1.29&  -6.98&22519.07&  7.87&  -0.25&  -7.44\\
 8693.95&  7.87&  -0.45&  -6.98&22552.57&  7.87&  -0.04&  -7.44\\
 8694.64&  7.87&   0.14&  -6.98&22563.83&  7.87&  -0.26&  -7.44\\
 9212.87&  6.53&   0.39&  -7.37&22575.39&  7.87&  -0.73&  -7.44\\
 9228.09&  6.53&   0.25&  -7.37&22644.06&  7.87&  -0.34&  -7.44\\
 9237.54&  6.53&   0.02&  -7.37&22707.74&  7.87&   0.44&  -7.44\\
\hline
\end{tabular}
\end{table}

\section{COMPARISON OF NON-LTE CALCULATIONS WITH OBSERVED STELLAR SPECTRA}

\subsection{Late-Type Stars}
After refining the S\,I line parameters based on the solar spectrum, we tested 
the sulfur atom model against spectra of late-type stars. We selected 
well-studied stars with reliably determined fundamental parameters. For an 
accurate atomic model, lines from various multiplets should be described by 
close sulfur abundances across a broad range of stellar atmospheric parameters. 
The study included the following stars. Three stars with solar-like composition: 
hot dwarf Procyon, hot giant HD\,195295, and cool giant Pollux. Two metal-poor 
stars: Ceti and HD\,22879, as well as the metal-deficient star HD\,84937. Thus, 
the stars used in the analysis differ greatly from each other. The stars were 
selected with low rotation velocities to minimize profile distortion.

Fundamental parameters were taken from \cite{Jofre15} for all stars except 
HD\,195295, for which data were taken from \cite{Lyubimkov10}. The spectrum for 
HD\,195295 was obtained using the 2.7-m telescope spectrograph at McDonald 
Observatory \citep{Tull95} with a resolution R = 60\,000. The spectra of the 
remaining objects, obtained with the UVES spectrograph at the Paranal 
Observatory \citep{Tull95}, were taken from the ESO archive. They all have a 
spectral resolution of at least 75\,000 and a signal-to-noise ratio of at least 
200. The spectra were processed and the continuum was drawn using the Dech 
package \citep{Galazutdinov22}. The atmosphere models of the stars were 
calculated using the ATLAS9 software with the ODF from \cite{Meszaros12}. The 
parameters of the stars are given in Table~\ref{Stars}. It also contains the 
results of determining the sulfur abundance for the multiplets used in the 
analysis (one line for each multiplet) and the corresponding non-LTE 
corrections. The abundances obtained from the lines within the multiplet are 
nearly identical. The given error in the average abundance, reflecting the 
line-to-line scatter, shows a very small variation. The real error in 
determining the element’s abundance will be slightly higher, since it should 
include the effect of the inaccuracy of the fundamental parameters of the star.

\begin{table*}
\centering
\caption {Sulfur abundance in the stars} 
\label{Stars}
\medskip
\begin{tabular}{cccccc|ccccc|c}
\hline
&\multicolumn{5}{c}{Stellar parameters} & \multicolumn{5}{c}{Sulfur abundance (S/H)}&Nl\\
\hline
Star& T$_{eff}$& log\,g&[Fe/H]&V$_{t}$&V~sin~i&6052&6757&8694&9212&10455& \\
\hline
Procyon	& 6554& 3.99& 0.01& 1.66& 2.8& 7.10& 	7.11& 7.15& 7.13& 7.13& 12\\
HD195295& 6570& 2.32& 0.00& 3.60& 8.0& 7.02& 	7.07& 7.06& 7.08& 7.08& 10\\
Pollux	& 4858& 2.90& 0.13& 1.28& 2.0&	   & 	7.20& 7.25& 7.15& 7.10&  6\\
$\tau$ Ceti& 5414& 4.49&-0.49& 0.89& 0.4&  & 	6.74& 6.67& 6.67&     &  6\\
HD 22879& 5868& 4.27&-0.86& 1.05& 4.4& 6.56& 	6.56& 6.53& 6.50&     &  6\\
HD 84937& 6356& 4.15&-2.03& 1.39& 5.2&     & 	5.40& 5.36& 5.34&     &  5\\
$o$ Peg	& 9597& 3.80& 0.25& 1.98& 5.4& 7.48& 	7.53& 7.56& 7.62& 7.52& 11\\
$\theta$ Vir& 9600& 3.60& 0.15& 1.42& 0.5& 7.40& 7.46& 7.44& 7.44&    &  7\\
HD 24040& 5809& 4.12& 0.09& 1.00& 1.5& 7.23& 	7.21& 7.19& 7.17& 7.20& 12\\
HD 28005& 5802& 4.18& 0.21& 1.00& 3.5& 7.34& 	7.35& 7.34& 7.36& 7.36& 10\\
HD 34445& 5803& 4.06&-0.03& 1.00& 4.5& 7.23& 	7.25& 7.22& 7.18& 7.18& 12\\
HD 82943& 5917& 4.23& 0.13& 1.00& 4.0& 7.32& 	7.34& 7.32& 7.32& 7.32&  9\\
HD 87359& 5645& 4.40&-0.07& 1.00& 4.0& 7.14& 	7.18& 7.14& 7.14& 7.16& 10\\
\hline
\end{tabular}
\begin{tabular}{ccc|cccccc}
&\multicolumn{2}{c}{Average (S/H)}&\multicolumn{5}{c}{Non-LTE corrections, dex} &  \\
\hline
           &Opt. range&H-range&6052&6757&8694&9212&10455\\
\hline
Procyon	   & 7.12 $\pm$ 0.02  & 		&-0.02& 	-0.02& -0.06& -0.48& -0.51\\
HD195295   & 7.08 $\pm$ 0.04  & 		&-0.06& 	-0.08& -0.21& -0.81& -0.83\\
Pollux	   & 7.17 $\pm$ 0.05  & 		& & 	 0.00& -0.02& -0.30& -0.18        \\
$\tau$ Ceti& 6.71 $\pm$ 0.04& 		& & 	 0.00&  0.00& -0.13&              \\
HD 22879&  6.54 $\pm$ 0.02& 		& 0.00& 	 0.01&  0.00& -0.19&      \\
HD 84937&  5.35 $\pm$ 0.03& 		& & 	 0.00&  0.00& -0.14&              \\
$o$ Peg	&  7.55 $\pm$ 0.05  & 		&-0.03& 	-0.04& -0.07& -0.37& -0.34\\
$\theta$ Vir& 7.43 $\pm$ 0.03& 		&-0.02& 	-0.03& -0.06& -0.35&      \\
HD 24040&  7.19 $\pm$ 0.02  &7.26$\pm$0.04	& 0.00& 	 0.00&  0.00& -0.20& -0.15\\
HD 28005&  7.35 $\pm$ 0.01  &7.32$\pm$0.03	& 0.00& 	 0.00&  0.00& -0.21& -0.15\\
HD 34445&  7.21 $\pm$ 0.02  &7.23$\pm$0.03	& 0.00& 	 0.00&  0.00& -0.22& -0.16\\
HD 82943&  7.32 $\pm$ 0.01  &7.30$\pm$0.03	& 0.00& 	 0.00& -0.01& -0.22& -0.15\\
HD 87359&  7.16 $\pm$ 0.02  &7.15$\pm$0.01	& 0.00& 	 0.00&  0.00& -0.17& -0.11\\
\hline
\hline
\end{tabular}
\end{table*}

\begin{figure*}
\resizebox{\hsize}{!}{\includegraphics{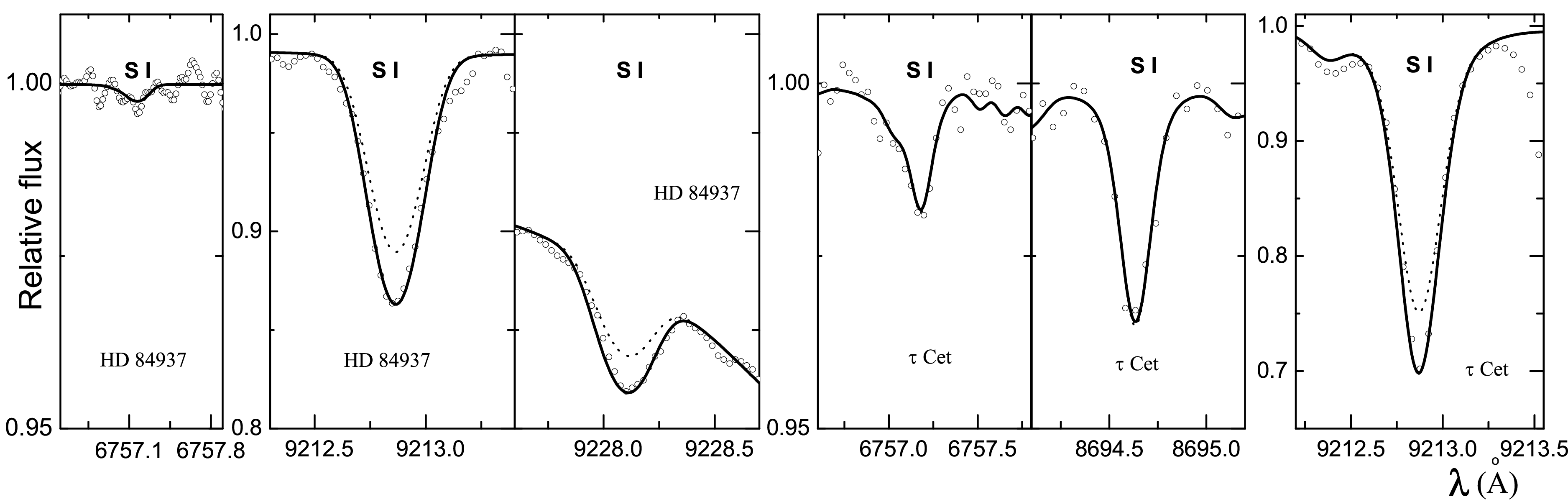}}
\caption{Comparison of observed (circles) and synthetic profiles of neutral 
sulfur lines in the spectra of HD\,84937 and t Cet. The non-LTE profile is 
shown as a solid line, while the LTE profile, calculated with the same 
abundance, is shown as a dashed line.}
\label{Cool}
\end{figure*}

The IR triplet at $\lambda$\,10455-10459~\AA, unfortunately, could only be used 
for Procyon, Pollux, and HD\,195295, since other spectra either had excessive 
noise in this range or did not cover it. For Pollux and $\tau$ Ceti, the 
$\lambda$\,6052~\AA\ line was heavily blended with CN molecular lines, and it 
was very weak for HD\,84937, so it was excluded from analysis. Lines of the 
eighth multiplet were also nearly invisible in HD\,84937, with equivalent 
widths below 1 m\AA. However, there is a clear depression at the 
$\lambda$\,6757~\AA\ line, which aligns well with the sulfur abundance 
determined from the first and sixth multiplet lines. This serves as an 
additional support for the derived sulfur abundance. Figure~\ref{Cool} shows 
examples comparing synthetic and observed profiles. It can be seen that, 
despite the varying sensitivity of lines from different multiplets to non-LTE 
effects depending on stellar atmospheric parameters, they yield closely aligned 
sulfur abundances. This can be considered as a good confirmation of the adequacy 
of the atomic model used in our analysis.

\subsection{Sulfur Lines in A-Type Stars}
The next step in validating our model involved applying it to hotter stars 
whose spectra feature sulfur in both S\,I and S\,II lines. These are A-type stars 
with temperatures above 9500\,K, where ionized sulfur lines become visible while
neutral sulfur lines in the visible spectrum also remain relatively intense. 
For this analysis, we selected two stars, $o$ Peg and $\theta$ Vir, studied in 
\cite{Romanovskaya23}. These stars have temperatures of 9600\,K and 
log\,g = 3.8 and 3.6, respectively. The spectra were acquired using the 
ESPaDOnS spectrograph \citep{Petit14} at a resolution R = 68\,000. For $o$ Peg,
a procedure of averaging 12 spectrograms was performed, which allowed the 
signal-to-noise ratio to be increased to 700. Details are given in 
\cite{Romanovskaya23}. 
 
\begin{figure*}
\resizebox{\hsize}{!}{\includegraphics{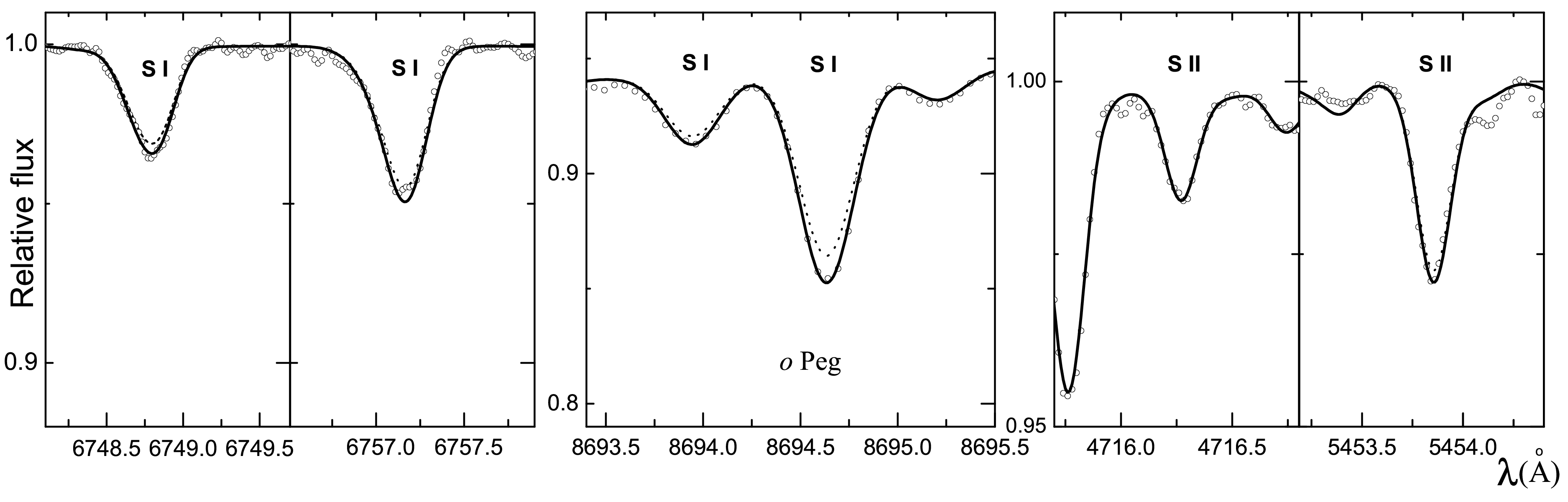}}
\caption{Comparison of observed (circles) and synthetic profiles of neutral and 
ionized sulfur lines in the spectrum of $o$ Peg. The non-LTE profile is shown 
as a solid line, and the LTE profile, calculated with the same abundance, is 
shown as a dashed line.}
\label{o_Peg}
\end{figure*}

The parameters of the ionized sulfur lines used for the analysis are given in 
Table ~\ref{lineSII}. The oscillator strengths are taken from \cite{Irimia05}. 
All lines are weak, from 2 to 7 m\AA, but are well detected due to the high 
quality of the spectra. These lines form deep in the photosphere and are 
therefore nearly free from the influence of non-LTE effects, which lead to a 
very small (by several percent) enhancement. Figure ~\ref{o_Peg} shows examples 
of comparison of theoretical and observed sulfur line profiles in two degrees of
ionization.

\begin{table}
\caption {Parameters of S II lines} 
\label{lineSII}
\medskip
\begin{tabular}{rcc|rcc}
\hline
$\lambda$ (\AA)& E$_{low}$& log~gf&$\lambda$ (\AA)& E$_{low}$& log~gf\\
\hline
4162.665&  15.944&  0.780 &5014.042&  14.067&  0.046 \\
4294.402&  16.135&  0.560 &5032.434&  13.671&  0.188 \\
4716.271&  13.617&  -0.365&5428.655&  13.584&  -0.177\\
4815.552&  13.671&  0.068 &5453.855&  13.671&  0.442 \\
4925.343&  13.584&  -0.206&5606.151&  13.733&  0.124 \\
5009.567&  13.617&  -0.234&5639.977&  14.067&  0.258 \\
\hline
\end{tabular}
\end{table}

Non-LTE results for neutral sulfur lines in the analyzed A-type stars are 
summarized in Table ~\ref{Stars}. For the S\,II lines, non-LTE sulfur 
abundances (S/H) $= 7.51 \pm 0.04$ fnd $7.38 \pm 0.04$ were obtained for $o$ Peg 
and $\theta$ Vir, respectively. These values show a difference of only 0.04 dex 
from the average sulfur abundance derived from S\,I lines. This is smaller than 
the uncertainties that may arise from oscillator strengths and fundamental 
stellar parameters. Thus, we conclude that the sulfur atom model is applicable 
up to effective temperatures of 10000\,K.

\section{NEUTRAL SULFUR LINES IN THE INFRARED H-RANGE}
Recently, spectral studies in the H-range have expanded. The emergence of 
high-resolution IR spectrographs, such as GIANO \citep{Origlia14}, has 
significantly enhanced the capability to determine the chemical composition of 
stellar atmospheres. This includes determining sulfur abundance based on lines 
from three infrared multiplets: $\lambda$\,15400-15422, $\lambda$\,15469-15478 
and $\lambda$\,22507-22707~\AA. In \cite{Korotin20}, some analysis of the 
non-LTE effects on these lines was performed using the 2009 sulfur model. The 
lines of two multiplets in the region of $\lambda$\,15400~\AA\ show minimal 
sensitivity to deviations from LTE, while the remaining lines form almost in 
LTE in the atmospheres of solar-type dwarfs. These conclusions remained 
unchanged with the use of a new atomic model. 
 
For testing, we used five stars from \cite{Korotin20}, which have IR spectra 
obtained with the GIANO spectrograph, as well as optical spectra from the ESO 
archive obtained with the UVES spectrograph. This allows a comparison of sulfur 
abundances determined from sulfur lines in the optical and H ranges. The 
parameters of the stars are given in Table~\ref{Stars}. The parameters of the 
S\,I lines used in the H-range are shown in Table~\ref{lineSI}. The oscillator 
strengths for lines of the two multiplets in the region of $\lambda$\,15400~\AA\
from BQZ and ZB have a systematic difference of 0.16 to 0.19 dex. As discussed 
in detail in \cite{Korotin20}, observed profiles for the Sun and the studied 
stars are better described when using the value $\rm log~gf = BQZ-0.12$ dex. 
For the $\lambda$\,22507-22707~\AA\ lines, the BQZ and ZB oscillator strengths 
differ randomly, yielding similar average abundances. The calculations used BQZ
data. The results in Table~\ref{Stars} show that the abundances obtained from 
lines in the optical range agree reasonably well with those from the H-range.

\section{NON-LTE CORRECTION GRID FOR NEUTRAL SULFUR LINES}

\begin{figure*}
\centering
\includegraphics[scale=0.35]{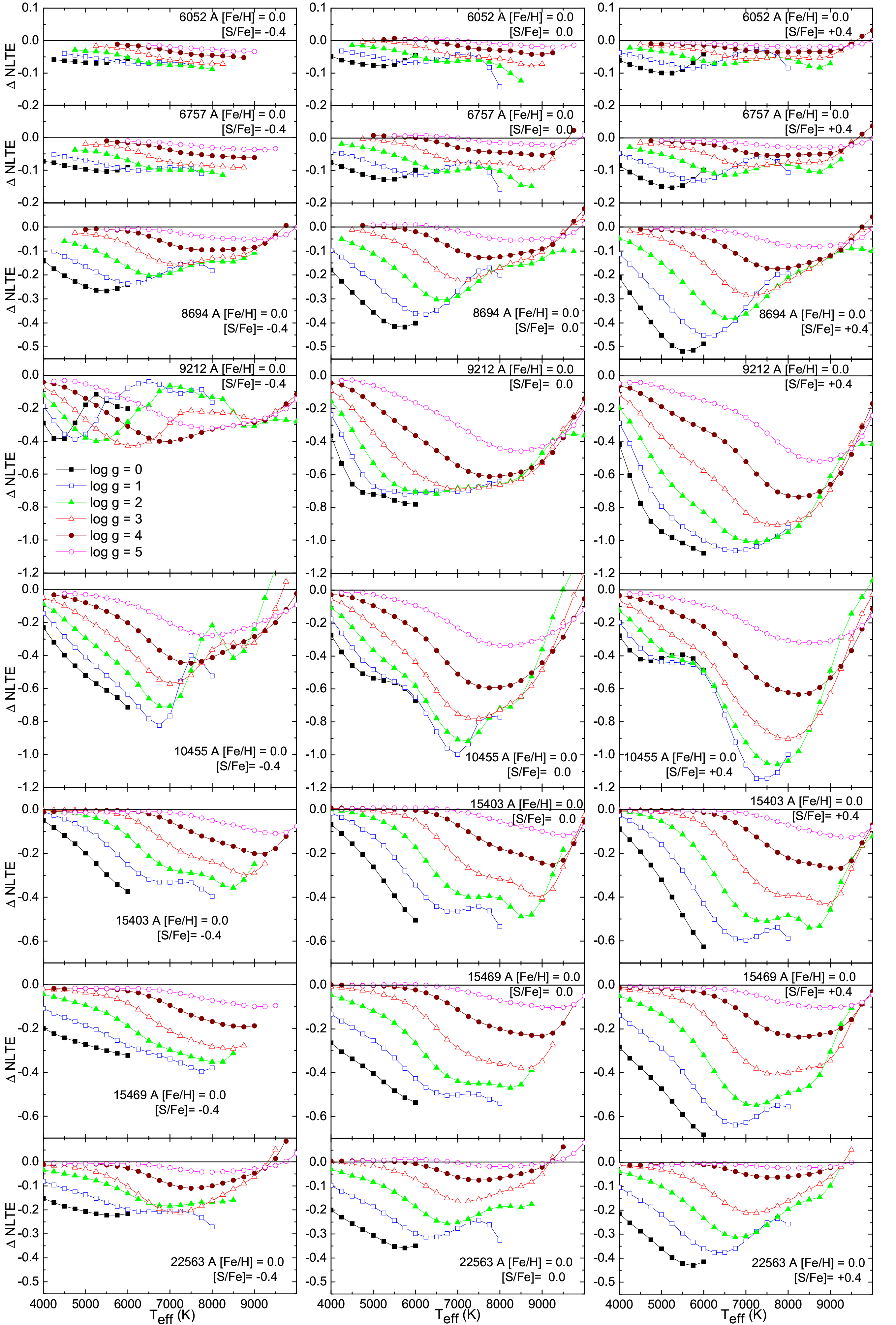}
\caption{Non-LTE corrections for metallicity [Fe/H] = 0}
\label{dNLTE}
\end{figure*}

\begin{figure*}
\centering
\includegraphics[scale=0.35]{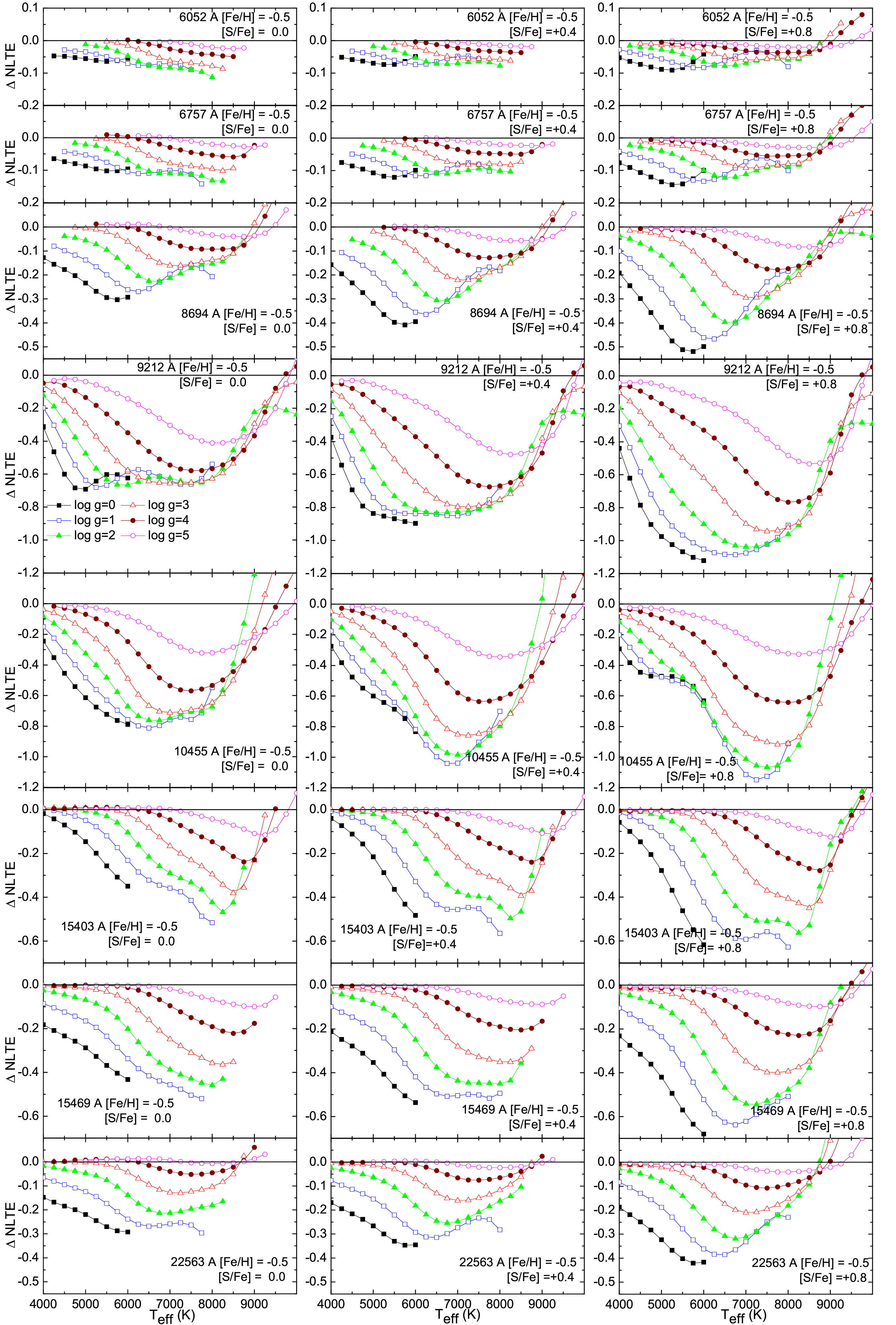}
\caption{Same as Fig.\ref{dNLTE} for [Fe/H]=-0.5}
\label{dNLTEm05}
\end{figure*}

\begin{figure*}
\centering
\includegraphics[scale=0.35]{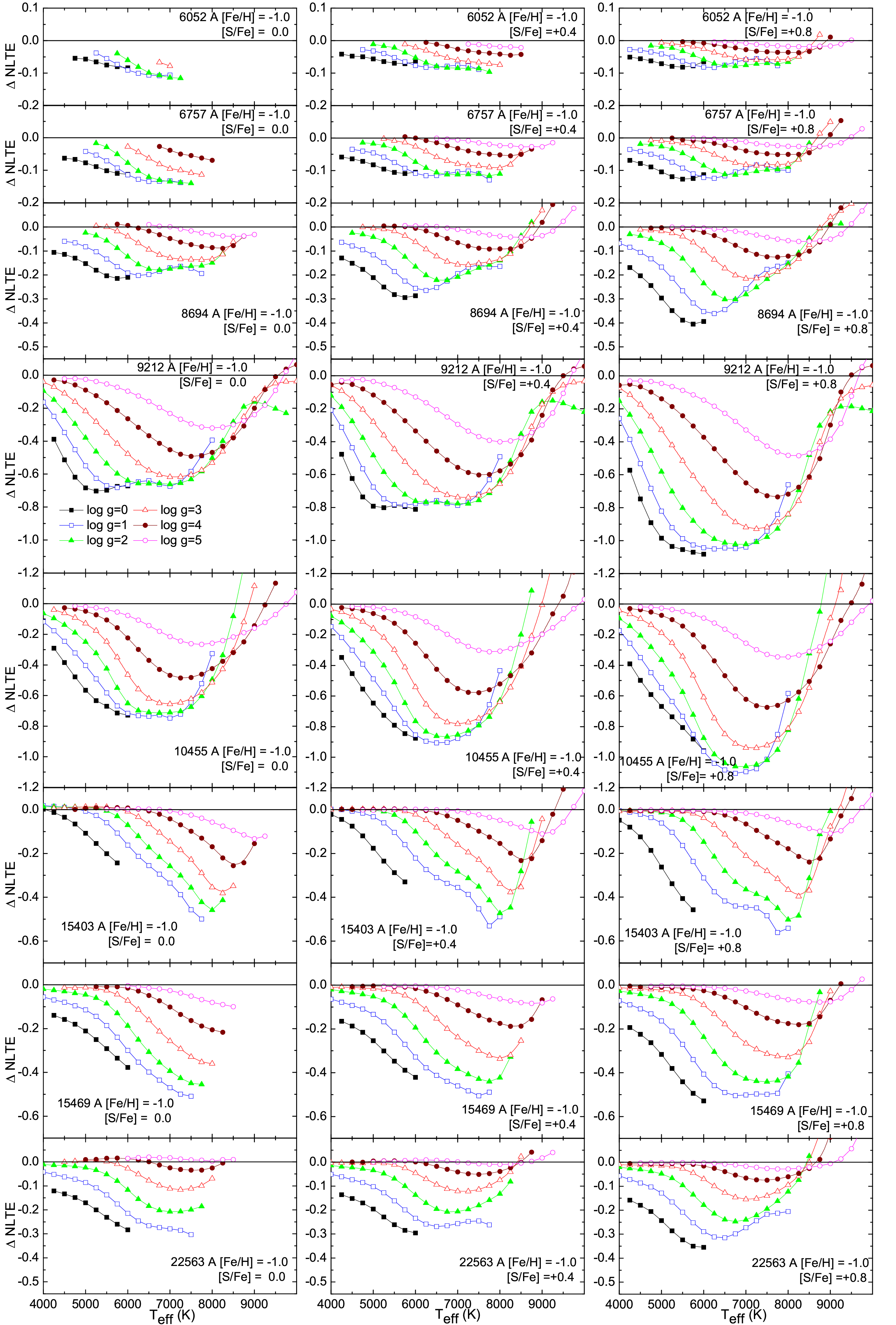}
\caption{Same as Fig.\ref{dNLTE} for [Fe/H]=-1.0}
\label{dNLTEm10}
\end{figure*}

\begin{figure*}
\centering
\includegraphics[scale=0.35]{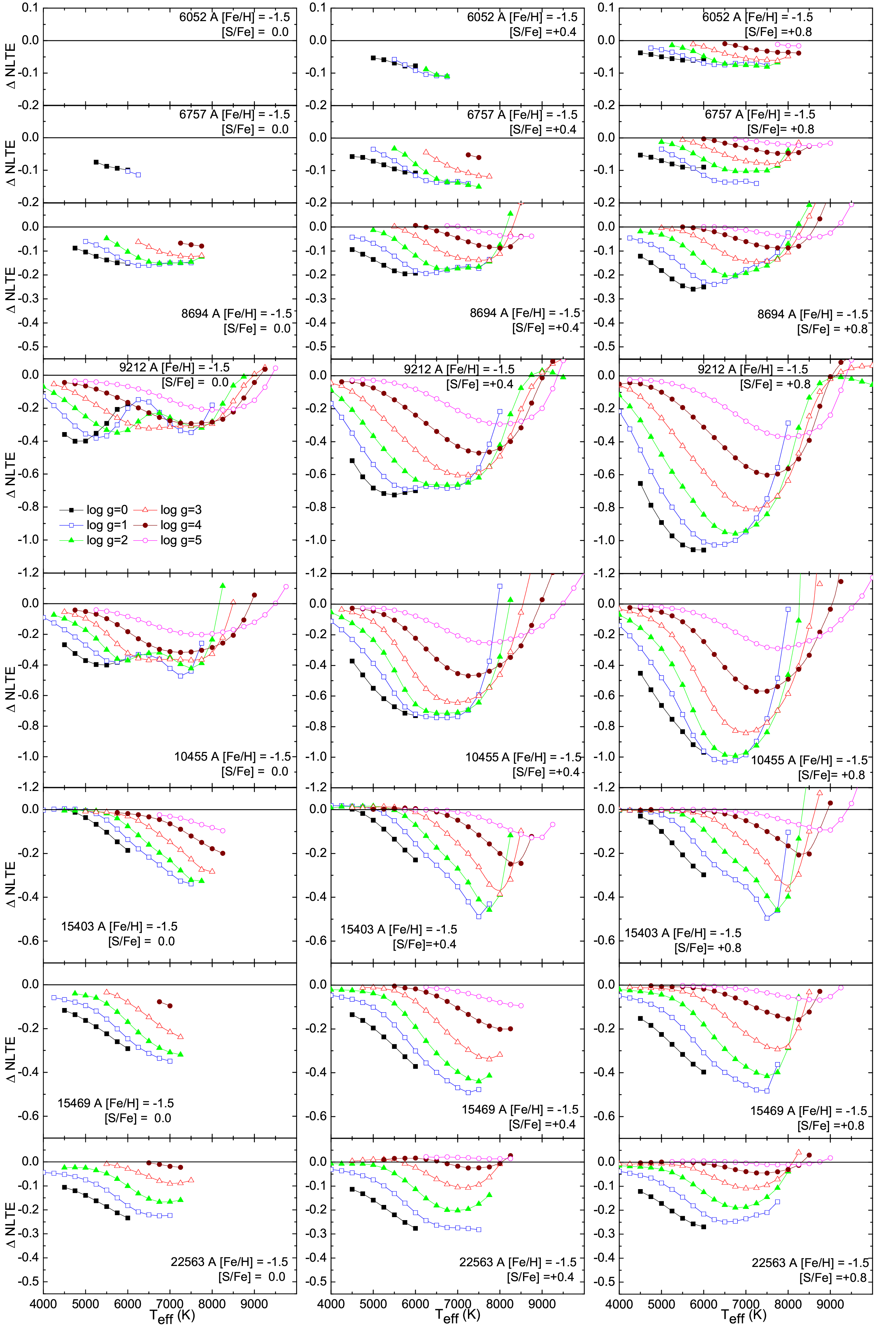}
\caption{Same as Fig.\ref{dNLTE} for [Fe/H]=-1.5}
\label{dNLTEm15}
\end{figure*}

\begin{figure*}
\centering
\includegraphics[scale=0.35]{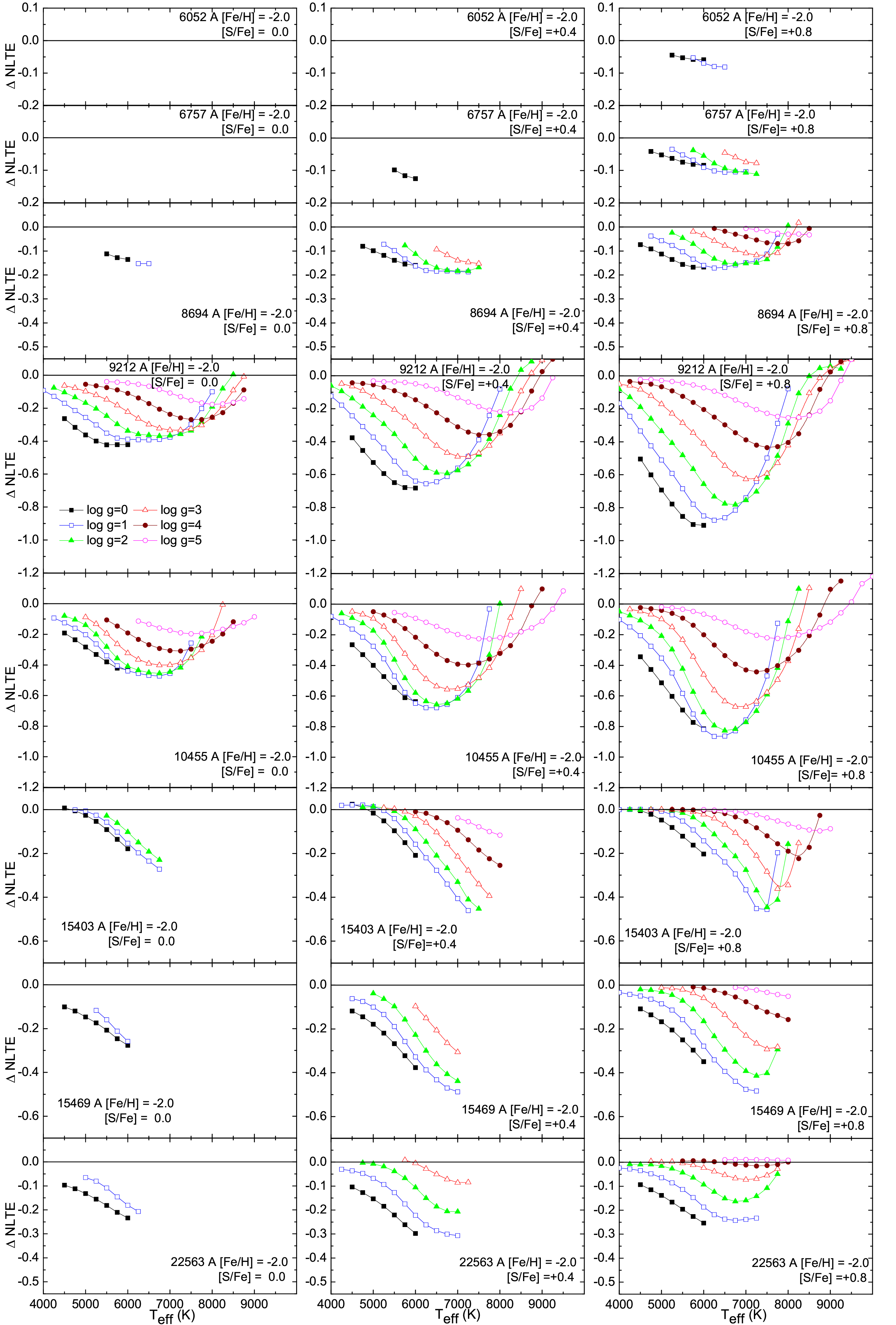}
\caption{Same as Fig.\ref{dNLTE} for [Fe/H]=-2.0}
\label{dNLTEm20}
\end{figure*}

After verifying the atomic model and refining the parameters of S\,I lines 
suitable for sulfur abundance determination, we calculated a grid of non-LTE 
corrections. Unlike the studies \cite{Takeda05} and \cite{Korotin09}, this grid 
includes lines in the H-range and extends to higher effective temperatures. The 
range of stellar atmosphere parameters spans effective temperatures from 4000 
to 10000\,K, acceleration of gravity log\,g from 0 to 5 and heavy element 
abundance [Fe/H] from 0 to –2. The microturbulent velocity was chosen to be 
2 km/s. Calculations for solar metallicity atmospheres were performed with 
sulfur abundances [S/Fe] = -0.4; 0.0; +0.4 dex, while for metal-poor models, 
the values were [S/Fe] = 0.0; +0.4; +0.8 dex, since such stars usually exhibit 
excesses of $\alpha$-elements. Calculations were performed by adjusting the 
sulfur abundance so that the equivalent width with non-LTE effects, matched the 
LTE value at the corresponding grid point. The difference between these 
abundances rep- resents the correction due to non-LTE effects. Corrections were 
not calculated for lines with an equivalent width less than 5 m\AA. It should 
be noted that the magnitude of non-LTE corrections can be significantly affected
by microturbulent velocity, since it can considerably alter the line formation 
depth for strong lines. The calculation results are presented in the form of 
graphs (Figs. \ref{dNLTE}-\ref{dNLTEm20}) since they visually illustrate and 
allow better assessment of the degree of non-LTE effects for each star, 
compared to the tables. Some general trends can be observed. As expected, the 
absolute value of corrections increases from dwarfs to giants, as the less 
dense atmosphere of giants reduces the impact of collisional processes that 
promote thermodynamic equilibrium. Lines of the eighth 
($\lambda$\,6743-6757~\AA) and tenth ($\lambda$\,6046-6052~\AA) multiplets in 
dwarfs have the smallest corrections and can be used in LTE analysis. Non-LTE 
effects on these lines are significant only for giants. The same conclusion 
applies to all lines in the H-range. Lines of the sixth ($\lambda$\,8694~\AA) 
multiplet are slightly more sensitive to LTE deviations, while lines from the 
two IR triplets can only be used if non-LTE corrections are applied. The 
non-LTE correction grid is available in electronic form. An example of these 
data is provided in Table~\ref{grid}, where the first column lists the 
wavelength, followed by [Fe/H], [S/Fe], and $T\rm _{eff}$ , with non-LTE 
corrections for five log\,g values in subsequent columns.

\begin{table*}
\centering
\caption {Grid of non-LTE corrections for S~I lines} 
\label{grid}
\medskip
\begin{tabular}{c|ccc|rrrrrrrr}
\hline
$\lambda$, \AA&[Fe/H]&[S/Fe]&T$_{eff}$&log\,g=0&log\,g=1&log\,g=2&log\,g=3&log\,g=4&log\,g=5\\
\hline
  6052.7&    0.0&   -0.4&  4000&        &        &        &        &        &        \\
  6052.7&    0.0&   -0.4&  4250&   -0.06&        &        &        &        &        \\
  6052.7&    0.0&   -0.4&  4500&   -0.06&   -0.04&        &        &        &        \\
  6052.7&    0.0&   -0.4&  4750&   -0.06&   -0.04&   -0.03&        &        &        \\
  6052.7&    0.0&   -0.4&  5000&   -0.07&   -0.05&   -0.03&        &        &        \\
  6052.7&    0.0&   -0.4&  5250&   -0.07&   -0.05&   -0.03&   -0.02&        &        \\
  6052.7&    0.0&   -0.4&  5500&   -0.07&   -0.06&   -0.04&   -0.02&        &        \\
  6052.7&    0.0&   -0.4&  5750&   -0.06&   -0.07&   -0.04&   -0.02&   -0.01&        \\
  6052.7&    0.0&   -0.4&  6000&   -0.06&   -0.07&   -0.05&   -0.03&   -0.01&        \\
  6052.7&    0.0&   -0.4&  6250&        &   -0.07&   -0.06&   -0.03&   -0.02&        \\
  6052.7&    0.0&   -0.4&  6500&        &   -0.07&   -0.07&   -0.04&   -0.02&   -0.02\\
  6052.7&    0.0&   -0.4&  6750&        &   -0.07&   -0.07&   -0.05&   -0.03&   -0.01\\
  6052.7&    0.0&   -0.4&  7000&        &   -0.07&   -0.08&   -0.06&   -0.03&   -0.02\\
  6052.7&    0.0&   -0.4&  7250&        &   -0.07&   -0.08&   -0.06&   -0.04&   -0.02\\
  6052.7&    0.0&   -0.4&  7500&        &   -0.08&   -0.08&   -0.07&   -0.04&   -0.02\\
  6052.7&    0.0&   -0.4&  7750&        &        &   -0.08&   -0.07&   -0.04&   -0.03\\
  6052.7&    0.0&   -0.4&  8000&        &        &   -0.09&   -0.07&   -0.05&   -0.03\\
  6052.7&    0.0&   -0.4&  8250&        &        &        &   -0.07&   -0.05&   -0.03\\
  ...   &   ... & ...   & ...  & ...    & ...    & ...    & ...    & ...    & ...    \\
\hline
\end{tabular}
\end{table*}

In general, the absolute value of corrections initially increases with 
effective temperature, since the UV radiation field actively affects the 
redistribution of sulfur level populations. However, upon reaching a maximum 
around 7000–8000\,K, the corrections start to decrease as more sulfur atoms 
become ionized, sulfur lines weaken, and form deeper in the star’s atmosphere 
where collisional processes become more influential. For stars with effective 
temperatures above 9500\,K, non-LTE corrections for most lines in dwarf stars 
become minimal or may even change sign. However, sulfur lines in such stars 
also become very weak. The correction value is also highly dependent on sulfur 
abundance, as it affects the intensity of sulfur spectral lines and, 
consequently, the depth at which they form.

Applying non-LTE corrections to determine abundances is generally reliable when 
the absolute correction value is less than 0.20–0.25 dex. However, it should 
be understood that the greater the correction, the larger its potential error 
due to complex dependences on microturbulent velocity, sulfur abundance, and 
other factors. Overall, the computed correction grid can be used to assess the 
degree of non-LTE effects on sulfur lines. For lines with larger non-LTE 
corrections, individual calculations of both b-factors and synthetic spectra 
are advisable.

If we compare our non-LTE calculations with those based on the 2009 atomic 
model \citep{Korotin17}, the corrections for stars with near-solar metallicity 
are quite similar between the two models. For metal-poor stars, the new model 
generally shows slightly smaller departures from LTE. For instance, using the 
old atomic model, for the $\lambda$\,9212~\AA\ line for a star with parameters
[Fe/H] = -1.0, log\,g = 2, and $T\rm _{eff}=6000$~K the correction was 
approximately –0.85 dex, whereas the new model yields a smaller value: –0.75 
dex. For dwarfs with log\,g = 4, the difference is smaller: $–$0.39 and
-0.34, respectively. This is due to the use of detailed quantum-mechanical 
calculations of collisional rates in the new model instead of approximate 
formulas from model \cite{Korotin09}. It is worth noting that there can be 
substantial differences between our data and calculations from \cite{Takeda05}.
Specifically, for the $\lambda$\,9212~\AA\ line in a star with parameters 
[Fe/H] = -1.0 and $T\rm _{eff}=6000$~K, the calculations in \cite{Takeda05} 
yield corrections of –1.02 for a giant (log\,g = 2) and –0.26 dex for a dwarf 
(log\,g = 4).The study \cite{Takada-Hidai96} provides non-LTE correction 
estimates for lines of the sixth multiplet ($\lambda$\,8694~\AA) in A-stars 
with log\,g = 4 and $T\rm _{eff}$ from 8000 to 12000\,K. The correction values 
for temperatures below 9500\,K are similar to ours, but at higher temperatures,
they remain negative, albeit small in magnitude.

\section{CONCLUSIONS}
We have proposed a grid of correction to take into account the departures from 
LTE in neutral sulfur lines in the visible and infrared spectral regions, 
including the H-range. The grid was calculated using a sulfur atomic model 
incorporating the most up-to-date atomic data on collisional rates with 
electrons and hydrogen. Inclusion of ionized sulfur levels and transitions in 
the atomic model has allowed us to extend the grid’s range of effective stellar 
photospheric temperatures up to 10000\,K. The atomic model has proven its 
adequacy across a wide range of stellar parameters: sulfur lines in all test 
stars showed consistent element abundances regardless of non-LTE effects. The 
wavelengths and oscillator strengths for several multiplets were refined. A 
recommended list of S\,I lines for determining sulfur abundance has been 
compiled. The recommended sulfur line parameters represent a compromise between 
observed stellar spectra and theoretical calculations of wavelengths and 
oscillator strengths. The best option for obtaining precise line parameters 
would be laboratory measurements for all sulfur lines used in the analysis.
 
\section{SUPPLEMENTARY INFORMATION}
The online version contains supplementary material available at 
https://doi.org/10.1134/S1063772924700987.

\bibliographystyle{aa}
\bibliography{Korotin}

\end{document}